# Publish or Patent:

# Bibliometric evidence for empirical trade-offs

# in national funding strategies

R. D. Shelton[1] and Loet Leydesdorff [2]

**Abstract**

Multivariate linear regression models suggest a trade-off in allocations of national R&D investments. Government funding, and spending in the higher education sector, seem to encourage publications, whereas other components such as industrial funding, and spending in the business sector, encourage patenting. Our results help explain why the US trails the EU in publications, because of its focus on industrial funding—some 70% of its total R&D investment. Conversely, it also helps explain why the EU trails the US in patenting. Government funding is indicated as a negative incentive to high-quality patenting. The models here can also be used to predict an output indicator for a country, once the appropriate input indicator is known. This usually is done within a dataset for a single year, but the process can be extended to predict outputs a few years into the future, if reasonable forecasts can be made of the input indicators. We provide new forecasts about the further relationships of the US, the EU-27, and the PRC in the case of publishing. Models for individual countries may be more successful, however, than regression models whose parameters are averaged over a set of countries.

[1] WTEC, 4600 Fairfax Drive #104, Arlington, VA 22203 USA; shelton@wtec.org.
[2] Amsterdam School of Communications Research (ASCoR), University of Amsterdam, Kloveniersburgwal 48, 1012 CX Amsterdam, The Netherlands; loet@leydesdorff.net.

# Introduction

A nation's scientific establishment (innovation ecosystem) can be considered as an economic system that needs inputs of resources in order to produce outputs that contribute to national prosperity with exports, jobs, and quality of life generally. Inputs to and outputs from such a system can be measured using indicators. Indicators are related variables that are easier to measure. Both authors of this study have found in a series of previous publications that a strong relation between input and output in science systems can be identified when the data is aggregated at the national level.

For example, Leydesdorff (1990) regressed percentages of world share of publications in the *Science Citation Index* (SCI) as output with *Gross Expenditure for R&D* (GERD) data as an input. This latter data is provided by the *Main Science and Technology Indicators*, published twice a year by the Organization of Economic Cooperation and Development (OECD). He noted that the advanced OECD-countries follow a common pattern of spending more money per paper over the years and labeled this effect as an R&D inflator. The marginal cost for improving one's relative share increases each year even after correction for inflation by the OECD.

At this time, the UK was performing above the regression line, while Japan underperformed in terms of returns on investment. In later studies, the measurement was refined by focusing on HERD (Higher Education Expenditure for R&D; Leydesdorff & Gauthier, 1996, p. 432) or HERD combined with GOVERD (intramural governmental expenditure for R&D) as independent variables (Leydesdorff & Wagner, 2009a, p. 357; Zhou & Leydesdorff, 2006, p. 90).

Shelton (2008) also used GERD as the independent (input) variable. Using a similar analysis, but independently, both Shelton & Foland (2009) and Leydesdorff & Wagner (2009a, at p. 356) predicted that China would surpass the United States in terms of internationally published articles, reviews, proceedings papers, and letters—that is, the citable items of the SCI—in 2014 or shortly thereafter, if growth and decline were to continue according to the same patterns. However, we assessed the policy implications very differently (Shelton, 2008; Leydesdorff & Wagner, 2009b).

Perhaps, this difference of opinion reflected our differences in backgrounds and positions on both sides of the Atlantic Ocean. The long term decline of the US output in share of world publications despite increased R&D investment was called "the American Paradox" by Shelton (2008). On the other side of the ocean, one is more concerned about the "European Paradox:" that is, Europe seems more efficient in investing in R&D in terms of publications, but patenting (a proxy of innovation) has remained behind the levels of the US (Foland & Shelton, 2010).

Dosi *et al.* (2006) hypothesized that the European Paradox originated in a general weakness in the performance of European universities. Various authors in Europe called for institutional reforms (Gibbons *et al*., 1994). Already in 1994, however, the European Union (EU-15) surpassed the US in terms of output—measured as world share of publications in the *Science Citation Index*. With hindsight, this can partly be attributed to the German unification in 1990 (Leydesdorff, 2000). It took Germany a few years to obtain synergetic surplus value from adding the knowledge base of East Germany into the Federal Republic (Leydesdorff & Fritsch, 2006).

The EU has recently been joined by twelve countries from Eastern Europe. Leydesdorff & Wagner (2009a, p. 359) found that the Slovak Republic now has the lowest average price per paper among the EU member states (US $62.7K). Japan and Austria are at the other end of the spectrum with prices per paper larger than US $200K. Thus, it seems that the EU has imported a cheaper labor force into its science system and thus gained in its efforts toward a knowledge-based economy (Leydesdorff, 2006).

In this study, we join forces to address systematically questions such as (*i*) the input and output measures that can be used in (multivariate) regression equations, (*ii*) the differences between the two paradoxes, and (*iii*) the extent to which predictions can be made reliably on the basis of the insights thus obtained. Following Foland & Shelton (2010), we extend the univariate regression analysis of inputs versus publication outputs to multi-variate analysis, and to add patent outputs. Furthermore, we specify policy implications that follow from these insights about dependencies in the dynamics of the two competing systems at the macro-level (US and EU), and at national levels, including China and Japan.

For this purpose, we use the various OECD indicators, publication and patenting statistics as available either online or in publications (e.g., the *Science and Engineering Indicators* of the National Science Board of the USA; National Science Board, 2010), and combine this data using multi-variate regression analysis in models using one or two independent variables. Our approach will be a mixture of exploration using the rich data sources and hypotheses testing.

Regression analysis cannot prove that a particular input causes an output. However, we will find here that certain funding components are more strongly correlated with publications and patents than other components. All researchers know that funding indeed is a necessary resource that empowers research and permits its outputs. Thus, the findings strongly suggest connections that might be verified by other means. All input and output indicators are correlated among one another, because they all tend to increase with the size of the country, and with time. Thus, we proceed cautiously and limit the analysis to a few independent variables in order not to over-fit our data.

**Methods and Materials**

*a. Data sources*

Input indicators considered here include: (1) The number of researchers in a country; (2) gross domestic expenditures on research and development (GERD); (3) four GERD *funding* components: government, industry, abroad, and other; and (4) four GERD *spending* components: higher education expenditures on R&D (HERD), business expenditures on R&D (BERD), government expenditures on R&D (GOVERD) and expenditures of non-profit institutes (other than universities). This investment data at the national level is available from OECD (2010) for its 39 member countries and affiliates. The data is already normalized by the OECD for inflation and for purchasing power parity (ppp).

Output indicators considered here include: (1) The annual number of scientific papers added to the Science Citation Index, including the Social Science Citation Index, (SCI; National Science Board, 2010) and Scopus (2010); and (2) three kinds of patent indicators: patent *applications* to the USPTO, patent *applications* in the Patent Cooperation Treaty (PCT) system, and triadic patent annual *grants* (registered in all of these three: the USPTO, the EPO, and the Japanese Patent Office). Most publication data is from the National Science Board (2010), based on fractional counts in the SCI. Scopus and the Web of Science were also used through their Web interfaces for more recent counts of publications.

Some data will not be included in the regressions. While a series is available for the EU-27, it is not used in regressions because most of these 27 countries are included individually. (Once regression equations are derived, the EU values can be calculated from them.) Some countries are omitted as outliers; e.g. the US is not included in the patent data from USPTO because the "home advantage" effect makes its data point atypical (Criscuolo, 2006).

Indicators may either be absolute numbers or relative, that is, percentage shares obtained by dividing by totals for a set of nations included into OECD (2010), which henceforth we shall call the "OECD+ Group." This is the set of 39 nations that have a fairly complete set of data from the Organization for Economic Cooperation and Development over many years (OECD, 2010).

Instead of absolute numbers, percentage shares often provide more revealing values for national comparisons, particularly when one wishes to compare results for two or more time samples. Papers published and patents granted are nearly a zero-sum game, because slots in journals, and the number of examiners available to process patents, do not rapidly change over time. Most of these indicators slowly increase with time, and this rising tide tends to raise all boats, obscuring national comparisons. For example, Thomson Reuters adds new journals each year to the SCI, making the database increase by some 3% per year (cf. Leydesdorff &Wagner, 2009b, Fig. 4, at p. 28). Sometimes, when a nation increases its total number of publications in this database by 3% in a year, or even less, some point with pride at their nation's performance. However, this can be just an artifact of the database.

Other output indicators like patents do much the same. Thus, in analyzing and modeling relative positions of nations, their share of outputs is a more relevant indicator than the absolute totals. However, this reasoning also applies to input indicators; since outputs are measured in terms of percentages of a total, modeling of the inputs that cause these outputs is also best done in terms of shares. This helps remove the effect of growth in time of all indicators. Of course, once a model is built for shares, it can easily be used to calculate absolutes, such as the total number of a nation's papers in the SCI.

The size of nations is another confounding effect. All inputs and outputs depend partly on the size of the country, which makes most country-wise correlations very high, obscuring identification of which variables are most important. One can divide all variables by some measure of size (e.g., population; cf. Dosi *et al*., 2006), but stepwise regression can tease out which inputs are best for predicting outputs. Independent variables (IVs) can be added one-by-one in order of which makes the best model for the prediction of the dependent variable (DV). The first IV absorbs the spurious correlation for size. .

Lags in time should also be considered. For example, one can expect that research funding takes some years to result in a scientific paper. There can also be a considerable lag between the time of patent applications and their grants, and even more from the initial R&D funding that enabled the invention. However, (except for China) most variables do not change rapidly from year to year, and strong correlations are obtained, even without lags. A further extension to ARIMA time-series models (e.g., in SPSS) is not pursued because we would lose the multivariate perspective and the advantage of stepwise introduction of the independent variables (cf. Leydesdorff, 1990).

Once a resource input has been identified that accurately predicts an output in the past it might be used to formally forecast future values of that output. This requires that the models be stable over time; here we usually compare models for two years: 1999 and 2007. It also requires a forecast of the input variables, but one is sometimes able to forecast these resource variables more easily than outputs because countries tend to change investments in R&D fairly smoothly with time. Furthermore, national governments sometimes publish S&T plans that state their future intentions for such investments.

*b. Regression and Interpretation*

While causality cannot be asserted from the results of exploring correlations and regression models, we proceed on the assumption that the input resources that are most predictive of outputs can be further examined as the most effective investments. For example, a nation's paper share of the OECD+ Group can be predicted as the dependent variable (DV) with a regression line that accounts for more than 95% of the variation using a single independent variable (IV): government R&D investment ($p < 0.001$). A similar regression with two IVs, the government and industrial funding components, will show that the industrial component is not significant ($p > 0.05$) as another IV in comparison. In our opinion, it can then be suggested that government funding is more important than industrial funding in producing scientific papers. Such a conclusion has also policy implications.

By regression over the 39 OECD+ countries, Shelton (2008) identified some of the national inputs that were most important, for example the number of researchers was not significant compared to investment. From the structure of the publication process, he built a model for individual countries in terms of shares:

$$m_i = k_i w_i \qquad (1)$$

In Equation 1, $m_i$ is the publication share, $k_i$ the relative efficiency, and $w_i$ the GERD share for the $i^{th}$ country. This model successfully accounted for the decline of the US and EU after 2000 as being due to China's rapidly increasing R&D investments. The relative efficiency $k_i$ happened to be fairly constant since 1998 for the US, EU, and PRC, permitting useful forecasts. As noted, Shelton and Foland (2010) used this model to forecast that China would soon pass the US and EU in papers in the SCI, as it already has in some physical science databases.

Thus, this model suggested that the GERD share has been the driver of changes in paper share, accounting for the rise of China since 2000 (Jin and Rousseau, 2005; Leydesdorff & Zhou, 2006; Moed, 2002; Zhou & Leydesdorff, 2006), and the inevitable associated (but relative) decline of the US and EU (Leydesdorff & Wagner, 2009b). In this individual country model, the relative efficiency does vary by country, and by time for some countries (Shelton and Foland, 2008).

Europe's very high $k_i$, however, remained a puzzle in Shelton's original model, and the model could also not account for the EU passing the US in the mid-1990s. More accurate models were needed that could account for Europe's rapid increase in efficiency during the 1990s (Foland and Shelton, 2010). Here, those models and similar ones for patents are presented in more detail. To permit comparisons of different models, they will be given names. Models for pape<u>r</u> shares will be named Mr1, Mr2, etc., while those for paten<u>t</u> share will be named Mt1, Mt2, etc.

**Results**

First, the method of stepwise inclusion of independent variables into the linear regression equation will be applied to the problem of predicting the shares of publications of nations, by searching for the IVs that best account for the dependent output variable (DV), that is, paper share in the SCI (including the SSCI; with fractional counts of articles, conference papers, and reviews; cf. National Science Board, 2010). Table 1 presents the correlations between SCI papers—the dependent variable—and the IVs considered to orient us in selecting the most useful models. The Scopus data (articles and reviews, for all fields) are presented only for comparison. The models will be based on the SCI paper shares.

**Table 1. Correlations for Papers ($N$ = 39 in the OECD+ Group)**

|  | SCI | | Scopus | |
|---|---|---|---|---|
|  | 1999 | 2007 | 1999 | 2007 |
| Capital vs. Labor |  |  |  |  |
|    GERD | 0.982 | 0.977 | 0.977 | 0.938 |
|    Researchers | 0.894 | 0.838 | 0.842 | 0.920 |
| Funding Components |  |  |  |  |
|    Industry | 0.973 | 0.959 | 0.968 | 0.920 |
|    Government | 0.989 | 0.989 | 0.986 | 0.944 |
|    Other | 0.917 | 0.948 | 0.909 | 0.924 |
|    Abroad | 0.672 | 0.657 | 0.565 | 0.592 |
| Spending Components |  |  |  |  |
|    HERD | 0.976 | 0.983 | 0.977 | 0.928 |
|    BERD | 0.980 | 0.968 | 0.975 | 0.927 |
|    Non-Profits | 0.925 | 0.975 | 0.914 | 0.951 |
|    Gov Labs | 0.985 | 0.961 | 0.984 | 0.938 |

All correlations are high because the paper output variables and all these input variables vary with the size of the respective countries. Thus, a similar model with a single IV could be constructed from any one of these measures. However, in a linear regression with two IVs, only the IV with the largest correlation will usually be found to be significant, since it also accounts for the underlying size factor when introducing the second IV.

For example the 2007 regression based on the capital (GERD) and labor (Researchers) components as IVs for SCI paper shares (as DV) is:

Mr1:   Papers07 = 0.819 GERD07 - 0.0270 Researchers07 + 0.536        ($R^2$ = 95.5%)

Clearly, the investment variable (GERD) is much more useful in predicting paper share than the number of researchers. Not only is the sign of the IV "Researchers" negative, but it is also not significant; its *p* value is larger than 0.7, far higher than the commonly used significance level of 0.05, while that of GERD parameter is significant ($p < 0.001$). (In the following models the significance level will be one permille ($p < 0.001$) unless stated otherwise.)

Since the number of researchers does not seem to add much precision, an overall regression model Mr2 could be built using GERD as a single input variable:

$$m_i = Kw_i + C \qquad (2)$$

While the model in Equation 1 had a relative efficiency, $k_i$, that differs by country, this model has a single average value for all countries, (capital) *K*, and it has a constant value *C* added. The new model enables us to specify when countries deviate from a common pattern among nations. The fit based on 2007 data is:

Mr2:  Papers07 = 0.800 GERD07 + 0.492 ($R^2 = 95.5\%$)

As in the case of Mr1, removing the insignificant Researchers IV from the model did not reduce the fit in terms of the $R^2$. Although this model has a good overall fit to the data ($R^2 = 0.955$), the paper shares for the US (29.9%) and the aggregated EU-27 (35.1%) are still far off the regression line, which predicts 28.3% and 20.1% respectively. However, the model seems to be rather stable over time; the same model using 1999 data is:

Mr2: Papers99 = 0.785 GERD99 + 0.564                                     ($R^2$ = 96.5%)

Table 1 and other multiple regressions indicate that the share of government *funding* part of GERD would be a more accurate predictor of papers. This model accounts for the EU increase in efficiency in the 1990s, but not yet for its passing of the US in the mid-90s. The regression equation for 2007 is:

Mr3:     Papers07 = 0.846 Government07 + 0.316                         ($R^2$ = 97.9%)

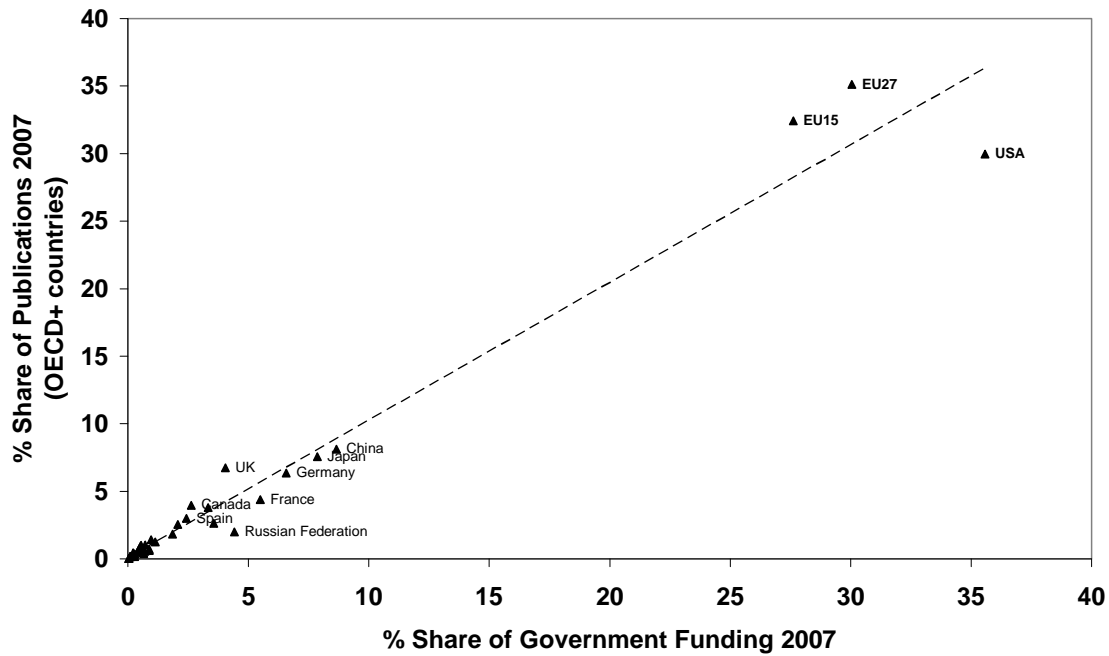

**Figure 1**: Paper share versus government funding normalized for OECD countries including the outlier values of the United States and the European Union.

Figure 1 shows this fit visually by drawing the regression line in Excel.[3] The fit seems partly an artifact from drawing a line between two clusters. In Figure 2, we focus on the smaller countries, that is, with less than 10% on both axes and including China and Japan: the fit is still high and the regression remains highly significant ($p < 0.001$).

---

[3] The parameters of the equation are different because the EU-15 and EU-27 are included in this representation, while these were considered as an aggregate of individual nations in the computations using MiniTab and SPSS.

As compared with previous analyses (Leydesdorff, 1990), Japan is placed in Figure 2 precisely on the regression line whereas the UK has maintained its outstanding position. The Russian Federation is relatively underrepresented in the international literature given its spending on R&D. Perhaps, one could consider the vertical deviation from the regression line as a measure of internationalization of national science systems in terms of (Anglo-Saxon) publication behavior.

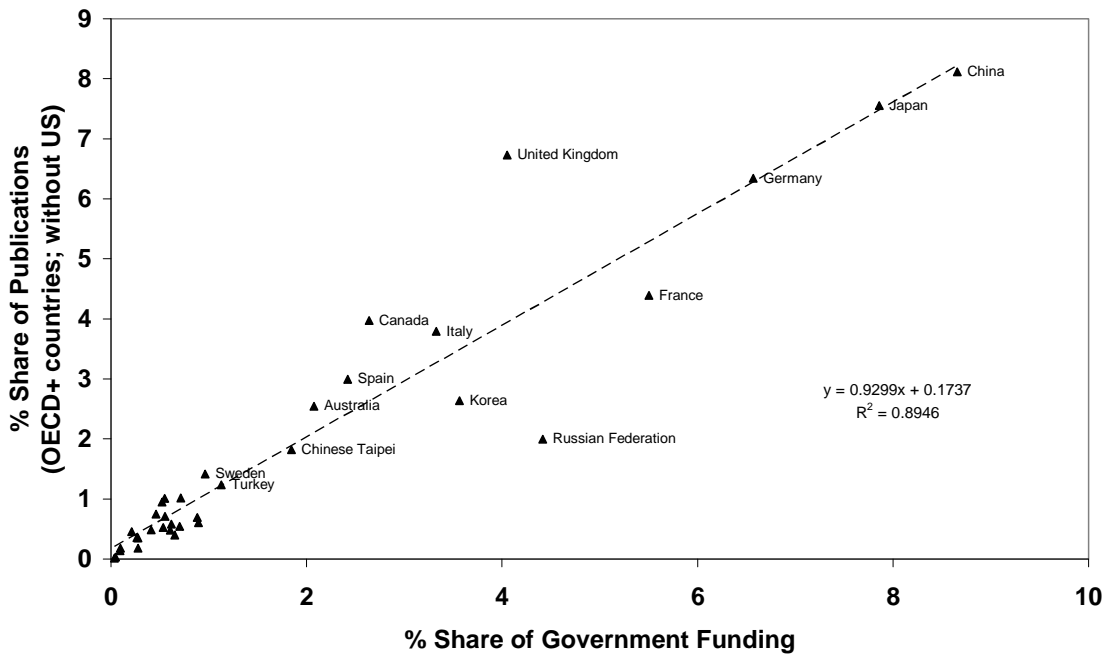

**Figure 2**: Paper share versus government funding normalized for OECD countries minus the outlier values of the United States and the EU.

Figure 1 shows these values including data for the USA and the EU. Since the US has a greater government share: 35.6% vs. 30.1%, the predicted US share is still larger at 30.4%, than the EU at 25.8%, The actual paper shares in 2007, however, were 29.9% and 35.1% respectively. Although these models are rather stable over time,[4] adding a second independent variable, the higher education spending share HERD, produces a more accurate model than Mr3. Specifically, the model for 2007 is:

Mr4:    Papers07 = 0.527 Government + 0.383 HERD + 0.127    ($R^2$= 98.8%)

Model Mr4 predicts 30.2% paper share for the US and 32.9% for EU-27, much closer to the actual values of 29.9% and 35.1% respectively, and demonstrates why it is reasonable that the EU should lead the US.

For comparison the same model with 1999 data is very similar:

Mr4:    Papers99 = 0.582 Government + 0.271 HERD + 0.249    ($R^2$= 98.2%)

An interesting, but simpler model based only on the HERD component as a single IV is successful in also predicting EU leadership in paper shares in 2007:

Mr5:   Papers07 = 0.979 HERD – 0.048    ($R^2$= 96.6%)

This model predicts 28.8% share for the US and 35.4% for the EU-27 in 2007; against the actual values of 29.9% and 35.1%, respectively.

---

[4] For comparison the same regression 1999 data is again not much different:

Mr2    Papers99 = 0.795 Government + 0.376    ($R^2$ = 97.8%)

**Table 2. Summary of Paper07 Regressions**

|     |   | IV1 | IV2 | Coeff1 | Coeff2 | Constant | p1 | p2 | $R^2$ |
|-----|---|-----|-----|--------|--------|----------|------|------|-------|
|     | 1 | Researchers |     | 0.789 |        | 0.521 | 0     |       | 70.2% |
| Mr1 | 2 | GERD | Researchers | 0.819 | -0.027 | 0.536 | 0.000 | 0.697 | 95.5% |
| Mr2 | 3 | GERD |     | 0.800 |        | 0.492 | 0.000 |       | 95.5% |
|     | 4 | Government | Industry | 0.774 | 0.067 | 0.330 | 0.000 | 0.351 | 97.9% |
| Mr3 | 5 | Government |     | 0.846 |        | 0.316 | 0.000 |       | 97.9% |
| Mr5 | 6 | HERD |     | 0.979 |        | -0.048 | 0.000 |      | 96.6% |
| Mr4 | 7 | Government | HERD | 0.527 | 0.383 | 0.127 | 0.000 | 0.000 | 98.8% |

Table 2 summarizes the models considered. In Row 1, one can build a single variable model using Researchers as the IV and produce a decent fit with $R^2 = 70.2\%$. However, Rows 2 and 3 show that GERD is a much more accurate predictor IV. Indeed, Researchers is not significant in comparison to GERD, and it adds no improvement to the fit (95.5%) over the single variable model in Row 3 (Mr2). In Row 4, the Industry funding contribution is not significant in comparison to Government funding. In Row 5 (Mr3), the best single variable model uses the Government share of GERD funding as the IV, although using HERD share works almost as well (Row 6, Mr5). The best two variable model in Row 7 (Mr4) uses the shares of Government funding and higher education spending (HERD) as IVs, and produces an excellent fit to the data (98.8%).

The model in Row 4 of Table 2 highlights that Government funding ($p < 0.001$) is more effective than Industry funding in generating papers in the SCI database.[5] This is hardly surprising, but regression results provide quantitative evidence for this intuitively expected result. However, one can draw several conclusions from the finding. One implication is that nations like the US that focus on industrial funding are likely to have lower publication outputs per GERD dollar than nations like the UK that focus more on government funding (Foland and Shelton, 2010).

---

[5] For both SCI and Scopus data government funding of R&D is better than industry funding in predicting the national number of papers. However, Table 1 shows that HERD and BERD have almost identical correlations with Scopus papers, so perhaps it is not so definite that higher education spending is better than business spending in producing papers. The inclusion of more trade journals in Scopus when compared with the ISI set may cause this difference. Furthermore, some preliminary results from the INSPEC database even show a correlation with government funding lower than that of the industrial funding component. This anomalous result probably comes from the narrow scope of INSPEC (mostly physical sciences only) compared to the much broader reach of the SCI and Scopus datasets.

**Patent Models**

Patents provide another output indicator of the performance of national research establishments. One complication is that most patents are awarded on a national rather than international basis. While some national offices, like the US Patent and Trademark Office (USPTO), receive many international applications, the United States itself dominates this dataset because of its "home court" advantage. However, the USPTO data is useful for evaluating the performance of other countries because the US market is attractive for foreign manufacturers, making US patents desirable internationally (Narin *et al*., 1997). Indeed, in 2008 foreign patents granted by the USPTO exceeded US grants for the first time. However, in regressions one normally removes such outliers as the US data point in this case, as being unrepresentative of the remainder of the data.

Two international patent series are available from the OECD. Triadic patents are those granted in all three: USPTO, European Patent Office, and the Japanese Patent Office. The second is patent applications under the Patent Cooperation Treaty (PCT), which has the advantage that they are much more numerous than triadic patent grants. Triadic patents, on the one side, are considered by National Science Board (2010) as of "high value" because applications to three national offices are costly, so that only the most promising patents are pursued. On the other hand, the PCT procedure allows applicants to seek patent rights in a large number of countries by filing a single international application with a single patent office, and then enter the national stage in the desired countries at a later date. These numbers are usually about three times those in the triadic patent series, and it will turn out that more accurate models can be based on these PCT patents.

Table 3 shows the correlations with the IVs again, but now compared to these three series for patent shares. At the two bottom lines of this table correlations between the three patent series are shown. Most correlations are very high, suggesting that several of the input variables could be used as a reasonable predictor of the output variables. Among the IVs available, we focus here on those components that are most controllable by government policy: GERD, Researchers, the Government and Industrial funding components of GERD, and the HERD (higher education) and BERD (business) spending components of GERD. The other components are too different between countries to draw as useful conclusions.

Table 3. Correlations for Patents (Shares)

| | Triadic 1999 | Triadic 2007 | USPTO 1999 | USPTO 2007 | PCT 1999 | PCT 2007 |
|---|---|---|---|---|---|---|
| Capital vs. Labor | | | | | | |
|   GERD | 0.924 | 0.895 | 0.947 | 0.830 | 0.974 | 0.963 |
|   Researchers | 0.847 | 0.680 | 0.664 | 0.428 | 0.845 | 0.762 |
| Funding Components | | | | | | |
|   Industry | 0.934 | 0.913 | 0.970 | 0.861 | 0.969 | 0.969 |
|   Government | 0.881 | 0.818 | 0.834 | 0.628 | 0.984 | 0.920 |
|   Other | 0.940 | 0.900 | 0.930 | 0.902 | 0.882 | 0.939 |
|   Abroad | 0.171 | 0.191 | 0.161 | 0.117 | 0.439 | 0.315 |
| Spending Components | | | | | | |
|   HERD | 0.949 | 0.890 | 0.910 | 0.791 | 0.961 | 0.960 |
|   BERD | 0.921 | 0.905 | 0.966 | 0.852 | 0.977 | 0.966 |
|   Non-Profits | 0.922 | 0.827 | 0.921 | 0.907 | 0.904 | 0.929 |
|   Gov Labs | 0.907 | 0.790 | 0.864 | 0.520 | 0.975 | 0.891 |
| USPTO Applications (US removed) | 0.971 | 0.956 | | | 0.774 | 0.915 |
| Triadic Grants | | | 0.971 | 0.956 | 0.883 | 0.972 |

Again slight differences in the correlations for the components can lead to the one with the lower correlation being found to be insignificant in a multivariate regression model. For example in all cases the component "Researchers" as IV has a lower correlation with all the DVs than GERD as IV. When regressions are made with this pair of IVs, the "Researchers" one will again be found to be insignificant compared to the GERD variable. Thus between these two, GERD seems not only to account for the country size factor, but to better account for other national factors.

In all cases shown in Table 3, except one, Industry funding is more highly correlated with patenting than Government funding. In the 1999 data for PCT patent grants, the Government funding is slightly higher correlated with output than the Industry variable. Likewise in all cases except one, the BERD spending component is more highly correlated with patenting than the HERD one. In the 1999 data for Triadic patent grants, the HERD component is slightly higher correlated than the BERD one.

*a. Triadic Patents*

As noted, we found that the capital variable (GERD) is much more important than the labor variable (the number of researchers) in predicting the patent share.

Mt1    Patents07 = 1.34 GERD07 – 0.465 Researchers07 + 0.327          ($R^2$ = 83.3%)

While the parameter "Researchers" as IV is significant at the 5% level ($p$ = 0.014), the sign of the component is again negative. As the correlations in Table 3 suggest, a single variable model can be built using the Industry funding share of GERD:

Mt2:   Patents07 = 0.941 Industry07 + 0.058          ($R^2$ = 83.4%)

The regression equation shown in Figure 3 is different from the one in Mt2 because the EU-27 data are weighted into this regression. The low value of the US in this case is a bit unexpected, but perhaps American firms patent primarily domestically. The relatively high values of Japan and Germany are noteworthy. Figure 4—analogously to Figure 2—enlarges the lower-left quadrant of the figure in order to see the patterns among "smaller" nations. China now deviates considerably with a participation in this type of patents much lower than average, whereas the Korean datapoint is similar to those of other OECD member states.

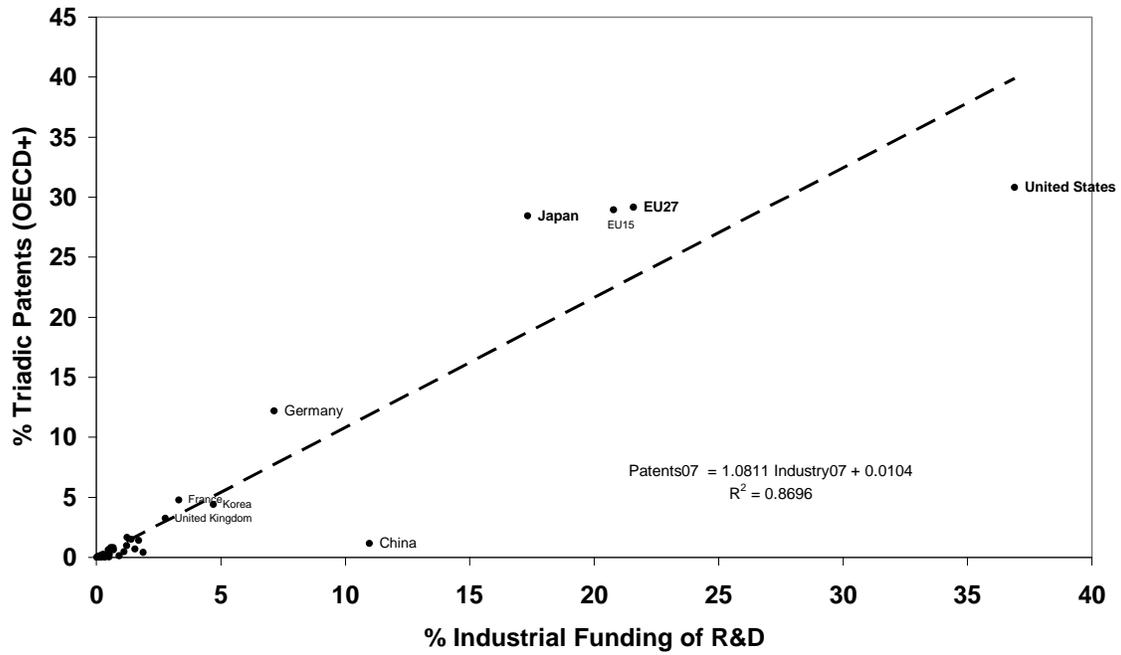

**Figure 3**: Triadic patenting versus Industrial Funding for the OECD+ set of nations.

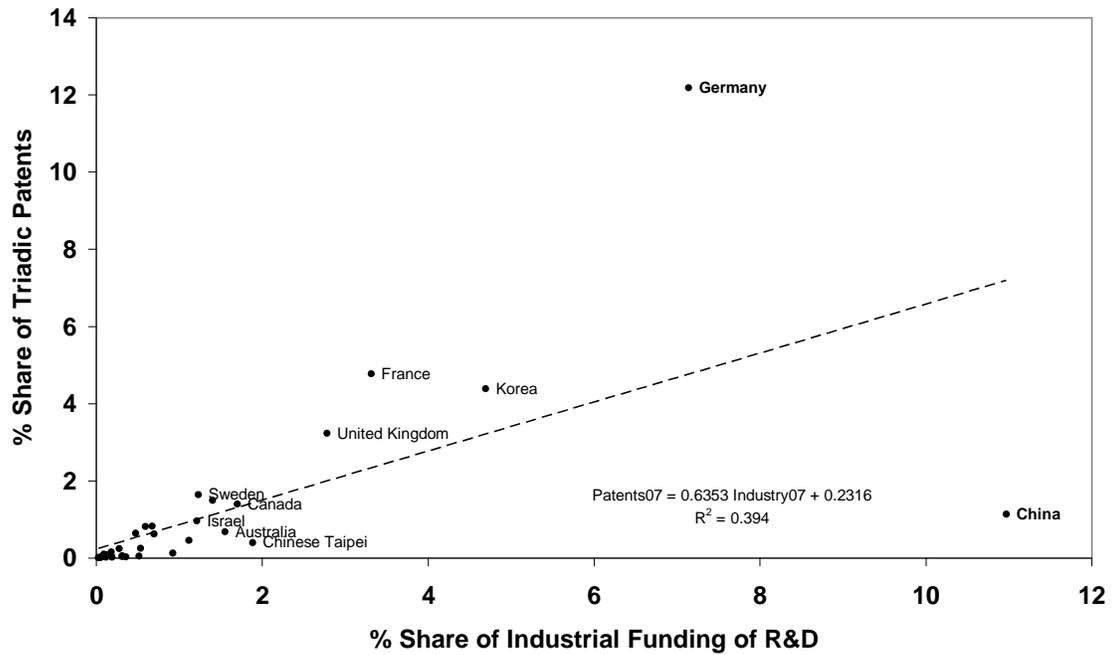

**Figure 4**: Triadic patenting versus Industrial Funding for "smaller" nations in the OECD+ set.

For comparison, a similar single IV regression with 1999 data has a somewhat more accurate fit:

Mt2:   Patents99 = 0.851 Industry99 + 0.298                         ($R^2$ = 87.2%)

Stepwise regression, however, suggested the following model to be considered:

Mt3:   Patents07 = 4.32 Industry07 – 3.46 BERD07 – 0.201            ($R^2$ = 85.9%)

The addition of the BERD component improves the fit of the regression line somewhat, but the negative sign is not so easy to explain. The negative signs suggest that for a given level of industrial funding, additional spending by businesses or nonprofits is not useful when producing patents is the objective—probably because the additional components focus on outputs other than patenting.[6]

Table 4. Summary of Patents07 Regressions (OECD+ set)

|  |  | IV1 | IV2 | Coeff1 | Coeff2 | Constant | p1 | p2 | $R^2$ |
|---|---|---|---|---|---|---|---|---|---|
| Mt1 | 1 | GERD | Researchers | 1.34 | -0.46 | 0.327 | 0.000 | 0.014 | 83.3% |
|  | 2 | Industry | Government | 1.78 | -0.973 | 0.438 | 0.000 | 0.000 | 88.6% |
| Mt 3 | 3 | Industry | BERD | 4.32 | 3.46 | 0.201 | 0.004 | 0.021 | 85.9% |
| Mt 4 | 4 | Industry | NonProfit | 2.04 | -0.653 | -0.584 | 0.000 | 0.000 | 98.3% |
|  | 5 | BERD | NonProfit | 2.28 | -0.771 | -0.828 | 0.000 | 0.000 | 97.3% |
| Mt2 | 6 | Industry |  | 0.941 |  | 0.058 | 0.000 |  | 83.4% |
|  | 7 | BERD |  | 0.953 |  | 0.078 | 0.000 |  | 81.8% |

---

[6] For comparison, a similar two IV regression with 1999 data has a better fit than in 2007:

Mt3:   Patents99 = 5.63 Industry99 – 4.82 BERD99 +0.477             ($R^2$ = 94.7%)

Table 4 summarizes the models for triadic patents in 2007. As in the case for papers, Row 1 again shows that the Researchers IV is not as important as GERD investment in predicting patents. Although the Researchers variable is now significant, its coefficient is still negative. In contrast to the papers case, however, Row 2 shows that the Industry funding component now is a much better predictor of patents than the Government funding component. Note that Government funding contributes negatively to patenting, and this relationship is significant ($p < 0.001$).

Row 4 (Mt4) is the best two variables model, with an excellent fit of 98.3%. However, these two IVs yield very different results using 1999 data, so this model does not seem to be stable over time. The best single IV model is in Row 6, using just the Industry funding component as an IV. Rows 6 and 7 show that Industry funding component is a somewhat more accurate predictor than the closely related business spending component (BERD)

*b. USPTO Patents*

Table 1 shows that USPTO patents are strongly correlated with the much less numerous triadic patents. Thus, models of triadic patents might suffice, but a couple of models for the USPTO are presented here.

Based on the USPTO application data in the OECD report, a fairly accurate single variable model can be built from the industry funding component of GERD. The "Other" funding source has a slightly higher correlation, but the nature of this funding from other national sources differs widely among countries, and is not useful for drawing conclusions. For 2007 the regression equation for share of US patents (with the US itself omitted) is:

Mt5:   USPTO07 = 0.771 Industry07 - 0.108                    ($R^2$ = 74.1%)

In this case, stepwise regression suggests a model with two independent variables:

Mt6:   USPTO07 = 1.29 Industry07 - 0.529 Researchers07 + 0.137    ($R^2$ = 92.9%)

As in the previous cases, the Researcher IV has a negative coefficient.

For reasons to be explained below, we also ran the following models with two independent variables:

Mt7:   USPTO07 = 1.120 Industry07 – 0.616 Government07 + 0.397    (R2 = 79.3%)

Despite its lower fit, this model is significant at the 1% level for the negative component of Government funding. When the outlier of US data is added to this set (Mt 7a), this effect is significant at the 1‰ level, but as stated before, this model may be over-fitted.

**Table 5. Summary of USPTO07 Regressions (OECD+ set)**

|      |   | IV1      | IV2         | Coeff1 | Coeff2 | Constant | p1    | p2    | $R^2$ |
|------|---|----------|-------------|--------|--------|----------|-------|-------|-------|
| Mt5  | 1 | Industry |             | 0.771  |        | -0.108   | 0.000 |       | 74.1% |
| Mt6  | 2 | Industry | Researchers | 1.29   | -0.529 | 0.137    | 0.000 | 0.000 | 92.6% |
| Mt7  | 3 | Industry | Government  | 1.12   | -0.616 | 0.397    | 0.000 | 0.008 | 73.3% |
| Mt7a | 4 | Industry | Government  | 2.055  | -1.015 | -0.505   | 0.000 | 0.000 | 96.4% |

*c. PCT Patents*

The other series available from the OECD is the number of patent applications filed under the Patent Cooperation Treaty (PCT). Table 6 shows some of the results in a similar format as the tables above.

**Table 6 Summary of PCT Patent Regressions in 2007**

|      |   | IV1        | IV2         | Coeff1 | Coeff2 | Constant | p1    | p2    | $R^2$ |
|------|---|------------|-------------|--------|--------|----------|-------|-------|-------|
|      | 1 | GERD       | Researchers | 1.180  | -0.314 | 0.325    | 0.000 | 0.001 | 94.7% |
|      | 2 | Industry   | Government  | 1.040  | -0.178 | 0.275    | 0.000 | 0.274 | 94.1% |
|      | 3 | GERD       |             | 0.946  |        | 0.101    | 0.000 |       | 92.8% |
|      | 4 | Government |             | 0.944  |        | 0.048    | 0.000 |       | 84.7% |
| Mt8  | 5 | Industry   |             | 0.887  |        | 0.205    | 0.000 |       | 93.9% |
| Mt9  | 6 | BERD       |             | 0.904  |        | 0.210    | 0.000 |       | 93.4% |
| Mt10 | 7 | BERD       | NonProfit   | 1.350  | -0.256 | -0.149   | 0.000 | 0.000 | 99.2% |

In 2007 the best single IV model is Mt8 in Row 5. Using BERD (in Mt9, Row 6) is almost as accurate. The best model with two IVs, however, is Mt10 in Row 7. The same models are very similar when run with 1999 data.

In summary, PCT patents are based more than triadic patents or USPTO patents on other sources of funding or spending components than industry funding. The Government component still contributes negatively as a second IV to the equation in Row 2, but this negative contribution is no longer significant. Thus, government and funding other than industry (but included in BERD) may have a more stimulating effect on adding to this wider pool of less-competitive patents (Leydesdorff, 2008).

**Forecasts from Models**

Regression models are intended to provide the best fit for a set of data for each single year. Strictly speaking, they are intended to predict the DV from IVs only for a new datapoint from the same population, i.e., a new country with similar characteristics and for the same year—as Chile as recently been added to the OECD. However, one purpose of this analysis is to try to develop models that can be used to forecast future values of DVs, particularly for the leading nations in publications and patenting. If the parameters of regression models are fairly stable over time, and IVs can be predicted more easily than the DVs, regression models could have some utility for forecasting. It will be seen that these models have only limited value in forecasting indicators for individual countries, unless they are made more elaborate. Thus a different approach will be proposed.

Direct regression of data on the time dimension is not advised because of auto-correlation in the data. Models which take this auto-correlation into account are more complex ARIMA models or of a different nature (Leydesdorff, 1990). However, one can try to fit the time-series in Excel, and make predictions without trying to model the underlying mechanisms. Figure 5, for example, shows an extrapolation of the data similar to the one by Leydesdorff & Wagner (2009a, at p. 356) which made these authors state that China could outperform the US in 2014 if the same trends would continue. More recent data (2003-2010) indicate that the growth of China is no longer exponential, but linear. Therefore, we would prefer a linear fit and therefore be inclined to predict a postponement of this date to the end of this decade.

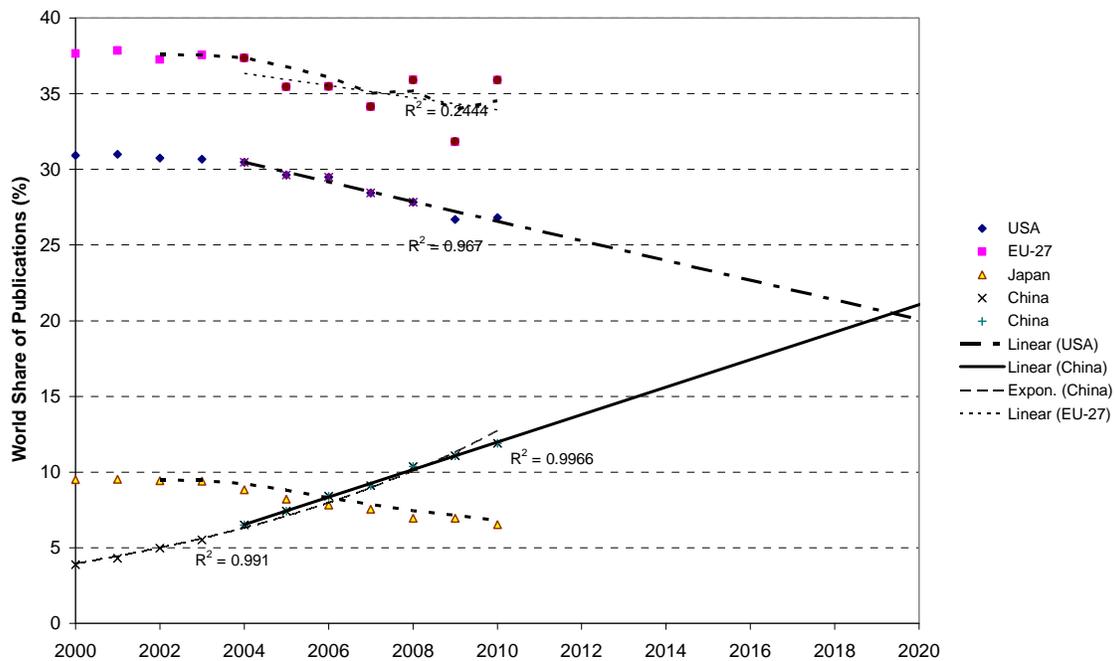

**Figure 5**: Forecast of percentage world share using extrapolation of a best fit based on data for 2003-2010 (cf. Leydesdorff & Wagner, 2009a, Figure 1, at p. 356).[7]

But to explore the feasibility of extended regression models, let us try to use 2005 models to predict 2007 as an example values for paper share for the three leading entities: the US, EU, and PRC, using Government funding as the single IV. The procedure used is to: (1) forecast this IV by extrapolating it from its 2005 value. The annual rate of change calculated over three years (2003-2005) is used as a slope; (2) The forecasted IV is then plugged into the model equation; (3) The model calculates a forecast for the DV for 2007. This can then be compared to the known actual value for 2007.

---

[7] Searches for 2010 were performed after the update of the database on January 22, 2011.

The model used for SCI paper share, based on 2005 OECD data as a function of the Government funding component normalized across the set, is:

Mr3:   Paper05 = 0.840 Gov05 + 0.662                                ($R^2$ = 96.5%)

**Table 7. Initial Values for the model in 2005**

|  | *Actual % of Papers* | *% Government funding* | *Model forecast of % of Papers* | *Error* |
|---|---|---|---|---|
| US | 30.95 | 37.4 | 32.04 | -1.09 |
| EU-27 | 35.41 | 30.4 | 26.20 | 9.21 |
| PRC | 6.27 | 6.90 | 6.46 | -0.63 |

Table 7 shows that using this model to forecast the future would start with a handicap in the initial year. The starting values for the US and PRC are fairly close, because their data points fall fairly near the regression line. However, the one for EU-27 is far off the line, perhaps because the EU as a whole was not in the dataset, only its individual member states. The EU is performing in terms of papers much better than expected.

Regression models are based on averages for a whole dataset. A prediction for an individual country may differ significantly from its actual subsequent value. The scattergrams in Figures 1 to 4 illustrate this. If a country's data point is far off the regression line in one year, this may be because of other factors that still persist in another year. Thus, an individual data point may well be far off the regression line for the second year, even if the regression lines for the two years are close together,

An example is in Table 8, which shows use of the 2005 regression line, with forecasted values of the IV (Government) to predict 2007 paper shares. These are compared to the actual data in the last two columns. Again the EU-27 value is far off, obviously because it started with a big offset. The other two forecasts are at least accurate in trends. They show the US declining and the PRC increasing, but both do so even more sharply than this model forecasts.

**Table 8. Paper forecasting parameters. Forecast is f; actual value is a.**

|  | *Gov05 (a)* | *Gov07 (f)* | *Gov07 (a)* | *Paper07 (f)* | *Paper07 (a)* |
|---|---|---|---|---|---|
| US | 34.4 | 37.3 | 35.4 | 32.0 | 30.0 |
| EU-27 | 30.4 | 29.0 | 29.8 | 25.0 | 35.1 |
| PRC | 6.9 | 7.9 | 8.6 | 7.3 | 8.1 |

While this is a negative result, the defects of the model offer several clues as to how more accurate models can be constructed. First one could enhance the model by adding the initial offset, so that at least the starting value is correct. Then the regression line's slope would more accurately forecast the changes that result when the input IV changes. The effect can easily be seen by adding the Error values in the right-most column of Table 7 to the forecasts in the Paper07 (f) column of Table 8. The new forecast values would be: 30.1, 34.2, and 6.7, respectively. This is much closer to the actual values of 30.0 and 35.1 for the US and EU. However, the difference between the lower new forecast for the PRC and the actual 2007 value of 8.1 actually increases.

Second, since the regression line is characterized by both an intercept and slope, one can consider another enhancement to change the slope to more closely fit the individual nation. However, this would bring us back to models for individual countries (Equation 1) or the curve-fitting exercise underlying Figure 5. One can produce more accurate forecasts at the expense of using individualized model parameters for the target countries, instead of a composite constant for the whole dataset. Individual country models at least initialize the forecast with the actual value for the starting year, removing one of the sources of error. Further, instead of using a composite slope for the regression line, a customized slope for that country can be used.

As a preliminary illustration of this method, Figure 7 shows some results for Model Mr1 in forecasting paper share from GERD share for three large entities: the US, PRC, and 15 countries of the EU-15. The parameters are based on 2005 data. To make for a more realistic test of the method, the GERD shares used are not the actual values, but rather 2005 forecasts based on earlier annual rates of its increase.

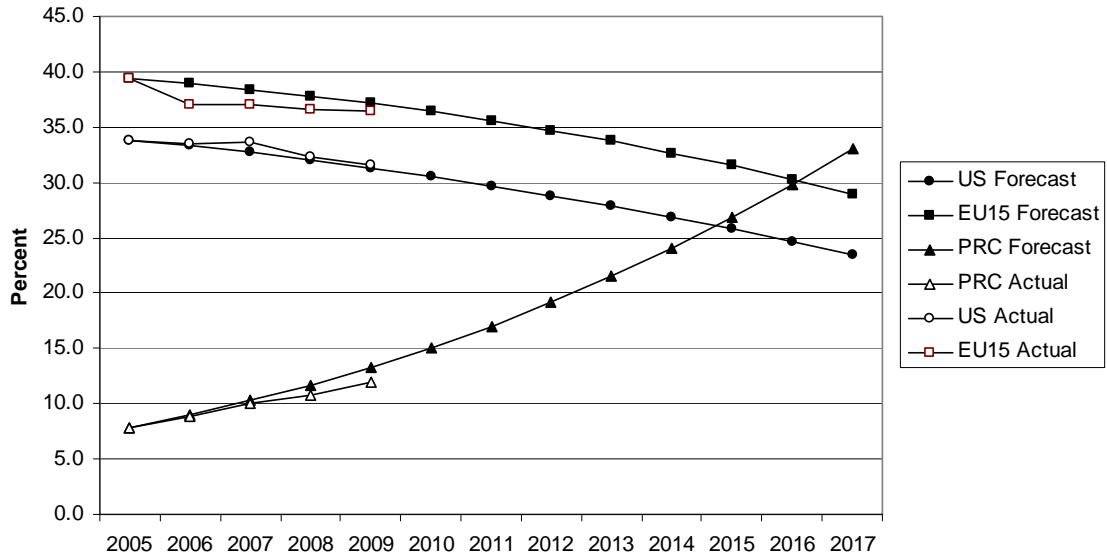

**Figure 6**: Accuracy of model Mr1 in forecasting paper share from GERD share. Note that the GERD share is forecast from the 2005 values.

These results are more useful. The main discrepancy shown is that the forecast value for the PRC is slightly below the actual values, which can be attributed to the fact that China's actual input GERD share has increased somewhat more than forecasted by extrapolation of IV trends before 2005. Using this forecast, the PRC would overtake the US some years earlier than using the extrapolations of Figure 5. Model Mr1 continues to be accurate for forecasts because the relative efficiency parameter ($k_i$) continues to be fairly flat for these three units of analysis (EU, USA, and PRC).

# Conclusions

The modeling process identified which resource IVs are most useful in predicting outputs, and shed some light on which investments are productive in enhancing these outputs. The analysis strongly suggests, although it cannot prove, that certain components of national R&D spending are more effective than others in producing papers and patents. The two components are complementary. That is, the Government funding and higher education spending is most effective in encouraging paper production. Industrial funding and business spending are conversely most effective in encouraging patent production. These findings are not surprising, but the regression models provide some quantitative evidence of their truth.

The model outcomes suggest that the Government funding component functions as a negative incentive to high-quality patenting. Because of the costs involved, corporations may be hesitant to invest in patents which were "not-invented-here," but these results can also be considered in the light of the continuous debate about the quality of university patents (Henderson *et al*., 1998; Leydesdorff & Meyer, 2010; Mowery & Ziedonis, 2002; Sampat *et al*., 2003). Perhaps, legislation such as the Bayh-Dole Act stimulates patenting for institutional reasons and hence may have undesirable institutional effects more than substantive contributions to industrial innovation (Mowery *et al*., 2003).

The models, furthermore, can shed light on the questions posed in the Introduction. The increases in publication efficiency in countries like the EU in the 1990s can be explained by their focus on the funding components that are most productive of papers. The American Paradox can be likewise explained since the Government funding component of GERD has been shown to be more important than overall GERD for the prediction of publishing. The US has a larger GERD than the next four largest nations combined, and has been steadily increasing this funding for 30 years. Despite this, it is now not surprising that its share of world paper production steadily decreased, because its share of the most decisive components Government investments and higher education spending have steadily decreased. Mainly this is because of the well-known shift in the proportions of government and industrial investment in the US: from about 50%: 50% in 1990 to about 30%:70% presently.

The European Paradox is the perception that the EU does not reap the full economic benefits of its leadership of scientific paper production. Our analysis also provides some insight into this observation. Although the US leads the EU in total R&D funding (GERD), the focus in the EU member states on the components that encourage paper production make it more reasonable that the EU should lead the US on this indicator. Conversely, the US focuses on the components that encourage patent production make it more reasonable that that it should lead the EU on this indicator. As shown in Figure 3, the EU is more efficient than the USA in obtaining patents from Industry funding. In summary, patents may be likely to lead more immediately to economic gains, but European research leading to papers represents a longer term investment.

Thus both the American and European Paradoxes seem to be the opposite sides of the same coin, reflecting complementary allocations of research investments. Furthermore, they can be interpreted as merely alternate choice between emphases on long-term research and more immediate development. The US seems to function as a more integrated system than the EU—with more functional differentiation between public and private. One should also keep in mind that the US is integrated as a national system, while the EU is not (Leydesdorff, 2000).

Let us finally note that regression models seem to be less useful in making forecasts for individual countries, since the composite slope and intercept of the regression line are based on averages over the dataset. The regression process is most useful in identifying the input IVs that are most useful for predicting output indicators, which then can be used to build models for the individual countries in question.

## Acknowledgements

Funding from NSF coop agreement ENG-0844639 is appreciated. The content is the opinion of the authors and not necessarily that of their institutions or funding agencies.